\documentclass{article}
\usepackage{spconfa4,amsmath,graphicx}

\usepackage[subtle]{savetrees}

\usepackage{hyperref}

\usepackage{cite}
\usepackage{amsmath,amssymb,amsfonts}
\usepackage{algorithmic}
\usepackage{graphicx}
\usepackage{textcomp}
\usepackage{xcolor}
\def\BibTeX{{\rm B\kern-.05em{\sc i\kern-.025em b}\kern-.08em
    T\kern-.1667em\lower.7ex\hbox{E}\kern-.125emX}}

\newcommand{\w}{\mathbf{w}}                         

\newcommand{\x}{\mathbf{x}}                         
\newcommand{\n}{\mathbf{n}}
\newcommand{\y}{\mathbf{y}}
\newcommand{\Rmat}{\mathbf{R}}  
\newcommand{\Rest}{\hat{\Rmat}}  
\newcommand{\Ry}{\Rmat_y}
\newcommand{\Rn}{\Rmat_n}
\newcommand{\Rx}{\Rmat_x}

\newcommand{\Rnest}{\Rest_n}
\newcommand{\Rxest}{\Rest_x}

\newcommand{\evec}{\mathbf{e}}
\newcommand{\refsub}{{\rm ref}}
\newcommand{\eref}{\evec_\refsub}
\newcommand{\Tr}[1]{\mathrm{Tr}\{#1\}}
\newcommand{\ypf}{\tilde{\y}}

\newcommand{\maskvec}{\mathbf{m}}              
\newcommand{\mask}{\mathcal{M}}                 

\newcommand{\expval}[1]{\mathcal{E}\{#1\}}
\newcommand{\inv}{^{-1}}

\title{A Study on Online Mask-based Beamforming Using Per-channel\\ Masking for Spatially Distributed Microphones}
\name{Wiebke Middelberg$^1$, Svantje Voit$^1$, Simon Doclo$^1$, Ryan M. Corey$^2$\thanks{This work was funded by the Deutsche Forschungsgemeinschaft (DFG, German Research Foundation) under Germany's Excellence Strategy - EXC 2177/2 - Project ID 390895286 and Project ID 568930428.}}
\address{$^1$ Dept. of Medical Physics and Acoustics and Cluster of Excellence Hearing4all,\\ Carl von Ossietzky Universität Oldenburg, Oldenburg, Germany\\ $^2$ University of Illinois Chicago and Discovery Partners Institute, Chicago, IL, USA\\ \{\href{mailto:wiebke.middelberg@uni-oldenburg.de}{wiebke.middelberg},\href{mailto:svantje.voit@uni-oldenburg.de}{svantje.voit}, \href{mailto:simon.doclo@uni-oldenburg.de}{simon.doclo}\}@uni-oldenburg.de, \href{mailto:corey1@uic.edu}{corey1@uic.edu}}
\begin{document}
\ninept
\maketitle
\begin{abstract}
Mask-based beamforming is a popular geometry-agnostic approach for speech enhancement, typically applying a single mask across all microphones to estimate the required covariance matrices. 
While effective for compact arrays, this strategy may be suboptimal for spatially distributed microphones, where signal characteristics may vary strongly across microphones. 
To effectively capture the spatial diversity across microphones, we extend the mask-based beamformer to a multi-channel formulation, where each microphone is pre-filtered by a separate mask before covariance estimation.
To address time-varying acoustic scenes, caused by spectro-temporal nonstationarity, we adopt a frame-causal online implementation with a sliding window. 
Experiments with simulated compact arrays and distributed microphones show that multi-channel masking yields a benefit over using a single mask when microphone signals differ substantially, while retaining similar performance in compact arrays. We further demonstrate the robustness of the multi-channel masking approach by comparing oracle ideal ratio masks to blind DNN-based mask estimation.
\end{abstract}
\vspace{-0.2cm}
\section{Introduction}
\label{sec:intro}
\vspace{-0.2cm}
In many speech communication tasks, noise reduction is a essential to allow for seamless inter-human and human-machine interaction. 
If multiple microphones are available, spatial or directional information can be exploited besides spectro-temporal information \cite{Doclo2015,Gannot2017}. 
In recent years, the interest in mask-based beamformers has increased \cite{Erdogan2016SingleMask,xiao2017mask,haebumbach2025ArrayDNN}, particularly for deep neural network (DNN)-based geometry-agnostic processing, where the number and position of microphones are unknown. In such a system, a mask is estimated -- for example with a DNN -- and applied to all microphone signals to roughly isolate the speech and noise components in each microphone. This can be interpreted as a per-channel pre-filtering. The pre-filtered signals are used
to estimate the spatial parameters for a signal-dependent beamformer, often a minimum-variance distortionless response (MVDR) beamformer, generating the enhanced output signal \cite{zhang2021AdHoc,jukic2023TAC,Tammen2024TAC,cornell2024chime,kamo2025ArrayIndep}.
This is particularly valuable for distributed or ad-hoc microphone arrays, which often have unknown or variable geometries \cite{zhang2021AdHoc,jukic2023TAC}. 
Typically, a single (average or median) mask is used for all microphone signals to estimate the speech and noise covariance matrix. 
The use of a single mask is a reasonable choice for compact arrays where all microphones typically capture signals with similar powers and signal-to-noise ratios (SNRs). However, in the case of spatially distributed microphones, signal characteristics (power, SNR, direct-to-reverberation ratio) may vary strongly across microphones \cite{Furnon2021tango,Corey2021,Middelberg_WASPAA23,Didier2024iDANSE}. In such a scenario, a single mask may be insufficient to optimally exploit the spatial diversity across microphones.

\begin{figure}
    \centering
    \includegraphics[width=0.99\linewidth,trim={3cm 22cm 1cm 3.5cm},clip]{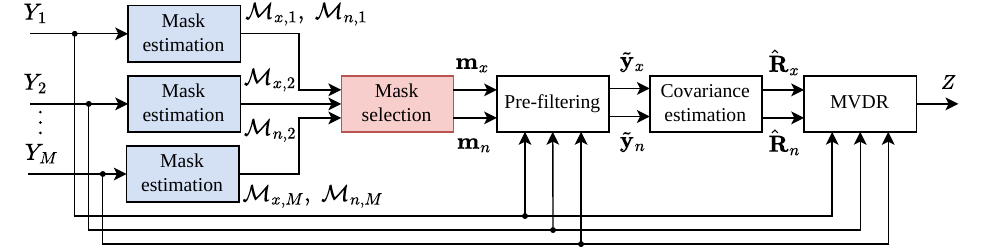}
    \vspace{-0.2cm}
    \caption{Block diagram for single- and multi-channel mask-based beamforming. The mask selection block determines the characteristics of the mask vector.}
    \vspace{-0.1cm}
    \label{fig:SigFlow}
\end{figure}

To improve the performance of mask-based beamforming for distributed microphones with high spatial diversity, in this paper we consider using multiple masks, i.e., applying a separate pre-filter to each microphone signal, shown in Fig. \ref{fig:SigFlow}.
The scope of this paper is the assessment of the benefits and limitations of multi-channel masking compared to single masks rather than techniques for mask estimation.
In \cite{Erdogan2016SingleMask}, a similar approach was proposed for the estimation of the noise covariance for compact arrays, where no significant benefit was observed. In \cite{wang2020spectralmapping,Ochiai2020beamtasnet}, a DNN-based per-channel pre-filtering stage was proposed to enhance the performance of an automatic speech recognition system. However, the proposed pre-filtering strategies were only tested on compact arrays and made use of task-specific single-channel filters and not general masks.
In this paper, we generalize the per-channel pre-filtering to a formulation of a multi-channel mask-based beamformer to account for the potential spatial diversity of distributed microphones. We compare three different masking approaches utilizing a single mask corresponding to the reference microphone, a mean mask across all microphones, and multi-channel masking, where an independent mask is applied to each microphone. 
Because distributed microphones are often used in complex sound scenes where source locations and activations change frequently, we consider a frame-by-frame online implementation. The adaptive beamformer is implemented using a sliding temporal window, for which we adopt a rank-2 covariance matrix update rule.

In the evaluation, we provide a comprehensive study on the benefits and limitations of multi-channel masking in an online implementation and investigate the effects of different temporal contexts. 
The results show little difference for compact arrays, but a benefit of multi-channel masking over using a mean mask for distributed microphones, especially in scenarios where the noise sources are mostly captured by different nearby microphones. 
To assess the robustness of multi-channel masking, we compare the performance using an ideal ratio mask and a DNN-based mask estimator and demonstrate that for a realistic mask estimator the multi-channel masking approach outperforms single masks, especially at low input signal-to-noise ratios.

\section{Signal Model}\label{sec:SignalModel}
\vspace{-0.2cm}
We consider a single target speaker in a noisy and reverberant environment, recorded by $M$ spatially distributed microphones with an arbitrary geometry. 
In the short-time Fourier transform (STFT) domain, the $m$-th microphone signal $Y_m(k,l)$ is given by
\begin{equation}\label{eq:ysum}
    Y_m(k,l) = X_m(k,l) + N_m(k,l)\, , \quad m\in\{1,\dots,M\}\, ,
\end{equation}
where $X_m(k,l)$ and $N_m(k,l)$ are the zero-mean target speech component and the noise component in the $m$-th microphone, respectively, and $k$ and $l$ denote the frequency bin and frame index. In the following, we will omit $k$ and $l$ wherever possible.
The noisy signal vector, $\y = [Y_1, Y_2,\dots,Y_M]^T$ with $\{\cdot\}^T$ the transpose operator, can be written as $\y = \x + \n$,
where $\x$ and $\n$ are the speech and noise vector, defined similarly as $\y$.
Assuming statistical independence between speech and noise, the noisy covariance matrix is given by
\begin{equation}\label{eq:Rydef}
    \Ry = \expval{\y\y^H} = \Rx + \Rn\, ,
\end{equation}
where $\Rx$ and $\Rn$ are the speech and noise covariance matrices, respectively. $\expval{\cdot}$ is the expectation operator, and $\{\cdot\}^H$ is the Hermitian transpose operator. 
The noise covariance matrix $\Rn$ is assumed to be full-rank but not following any particular spatial distribution. 
It should be noted that in a scenario with distributed microphones, different microphones might capture different noise sources. Hence, there may be a low correlation between the noise components in different microphones \cite{Middelberg_WASPAA23} and the noise power spectral density and SNR may vary  across microphones.

A popular geometry-agnostic signal-dependent spatial filter is the MVDR beamformer, which can be written in terms of the covariance matrices \cite{souden2010study} as
\begin{equation}\label{eq:w2}
    \w = \frac{\Rnest\inv\Rxest}{\Tr{\Rnest\inv\Rxest}}\eref\, ,
\end{equation}
where $\Tr{\cdot}$ denotes the trace of a matrix and, $\eref$ is a selection vector for the reference microphone containing all zeros except for the reference element. In this formulation, $\Rxest\eref$ can be interpreted as an estimate of the steering vector of the target speaker, representing the acoustic propagation between microphones.
The output signal $Z$ of the beamformer is given by $Z =  \w^H\y$.

In the following, we will focus on mask-based approaches for estimating the speech and noise covariance matrices $\Rest_\nu$ with $\nu \in \{x,n\}$, which are required to compute the MVDR beamformer in \eqref{eq:w2}. 

\section{Online Mask-Based Covariance Estimation}
\vspace{-0.2cm}
In this section, we present mask-based estimation methods for the speech and noise covariance matrices. We first generalize the mask-based estimator (using a single mask) to a multi-channel pre-filtering to account for spatial diversity, for which we consider different mask vectors based on either a single or multiple masks.
Lastly, we introduce a frame-causal online implementation to account for temporal changes caused by the intermittency of sources.

\vspace{-0.1cm}
\subsection{Covariance Matrix Estimation}\label{subsec:multimask}

A typical formulation for mask-based covariance estimation uses a single real-valued mask $\mask_\nu$ that is applied to all microphones per time-frequency bin and the average is calculated over a full utterance with $L$ frames \cite{jukic2023TAC,Tammen2024TAC,haebumbach2025ArrayDNN}, i.e.,
\begin{subequations}
    \begin{align}
    \Rest_\nu(k) &= 
\frac{1}{L} \sum\limits_{l=1}^{L} \mask_\nu(k,l)\, \y(k,l)\y^H(k,l)\label{eq:batchCovMatSingle}\\ 
&= \frac{1}{L} \sum\limits_{l=1}^{L} \ypf_\nu(k,l)\ypf_\nu^H(k,l)\, ,\label{eq:batchCovMatVec}
\end{align}
\end{subequations}
where $\ypf_\nu$ denotes the pre-filtered signal vector
\begin{equation}\label{eq:prefilter}
    \ypf_\nu= \sqrt{\mask_\nu}\, \y = \sqrt{\maskvec_\nu} \odot \y\, ,
\end{equation}
with $\odot$ the element-wise multiplication, and $\sqrt{\cdot}$ the element-wise square root.

While averaging over a full utterance, as in \eqref{eq:batchCovMatVec}, allows for estimating the signal statistics reliably in spatially and spectro-temporally stationary scenarios, it does not allow for causal real-time applications or tracking temporal changes in the acoustic scene.
Especially distributed microphones -- potentially capturing different, intermittent sources -- are subject to spectro-temporal nonstationarity of the present noise field.
To handle dynamic acoustic scenarios, we consider a frame-causal online approach to estimate the time-varying covariance matrices. Various approaches exist, e.g., using a sliding window \cite{Ochiai2023attentionmask}, recursive smoothing \cite{Higuchi2018AdaptMVDR,wang2020spectralmapping}, attention weights \cite{Ochiai2023attentionmask,bai2025attention}, or stochastic models \cite{Kubo2019timevarying}. In this paper, we opt for an online implementation using a short window, consisting of $P$ frames (temporal context), i.e.,
\begin{equation}\label{eq:onlineCovMat}
    \Rest_\nu(k,l) = 
\frac{1}{P} \sum\limits_{p=0}^{P - 1} \ypf_\nu(k,l-p)\;\ypf_\nu^H(k,l-p)\, .
\end{equation}

The pre-filtering operation in \eqref{eq:prefilter} aims at coarsely separating the speech and noise components in each microphone without distorting the inter-channel characteristics like phase and amplitude differences. 
In \eqref{eq:prefilter}, the same mask is used across all microphones, for which the mask vector is defined as
\begin{equation}\label{eq:maskvecsingle}
    \maskvec_\nu = [\mask_{\nu},\mask_{\nu},\dots,\mask_{\nu}]^T\, .
\end{equation}
However, the formulation in \eqref{eq:batchCovMatVec} can straightforwardly be extended to multiple masks by pre-filtering each microphone independently, similarly to \cite{Erdogan2016SingleMask}, e.g., in order to account for the spatial diversity expected for spatially distributed microphones.
To generalize the mask-based beamformer in \eqref{eq:batchCovMatVec}, we allow $\maskvec_\nu$ to take different values for each channel, i.e.,
\begin{equation}\label{eq:maskvec}
    \maskvec_\nu = [\mask_{\nu,1},\mask_{\nu,2},\dots,\mask_{\nu,M}]^T\, .
\end{equation}
In compact arrays, all masks can be expected to be similar as all microphones capture signals with approximately the same powers and SNRs. Hence, we expect \eqref{eq:maskvecsingle} and \eqref{eq:maskvec} to yield similar mask vectors. 
For distributed microphones, signals may have different relative powers and SNRs across microphones. Hence, there can be a benefit in applying different masks for each microphone, such that they contribute differently to estimating the speech and noise covariance matrix, respectively.

\subsection{Mask Selection}\label{subsec:mask}
In this paper, we will investigate different mask vectors in \eqref{eq:maskvec}, which is represented in the "Mask selection" block in Fig. \ref{fig:SigFlow}.
We consider three mask vectors, where "ref" and "mean" are single masks, and "multi" refers to an individual mask per microphone:

\textbf{Option 1 - reference mask for all microphones:}
Using the estimated mask corresponding to the reference microphone across all microphones with the mask vector
\begin{equation}\label{eq:maskvec1}
    \maskvec_\nu^{\rm ref} = [\mask_{\nu,\refsub},\mask_{\nu,\refsub},\dots,\mask_{\nu,\refsub}]^T\, .
\end{equation}
As a reference microphone, we select the one closest to the target speaker to provide a well-chosen mask, which in practice requires reference microphone selection.

\textbf{Option 2 - mean mask for all microphones:}
Using the average mask for all microphones, similarly to \cite{zhang2021AdHoc,Tammen2024TAC}, i.e.,
\begin{equation}\label{eq:maskvec2}
    \maskvec_\nu^{\rm mean} = [\overline{\mask}_\nu,\overline{\mask}_\nu,\dots,\overline{\mask}_\nu]^T\, ,
\end{equation}
where $\overline{\mask}_\nu = \frac{1}{M} \sum_{m=1}^M \mask_{\nu,m}$ is the average mask over all microphones.

\textbf{Option 3 - different mask per microphone:} Using the estimated mask \textit{per microphone}, corresponding to a multi-channel masking approach, with the mask vector
\begin{equation}\label{eq:maskvec3}
    \maskvec_\nu^{\rm multi} = [\mask_{\nu,1},\mask_{\nu,2},\dots,\mask_{\nu,M}]^T\, ,
\end{equation}
which allows microphones to contribute differently to the estimated covariance matrices.

\section{Evaluation}\label{sec:Eval}
In this section, we present the evaluation of the different masking approaches described in Section \ref{subsec:mask}. We conduct three different experiments investigating 1) the effect of the temporal context for the different mask vectors using distributed microphones, 2) the differences between performance using a compact array compared to distributed microphones, and 3) the robustness of the considered mask vectors at different input SNRs.

\vspace{-0.1cm}
\subsection{Acoustic Scene and Data Generation}\label{subsec:Data}

\begin{figure}
    \centering
    \includegraphics[width=0.93\linewidth,trim={0cm 0.2cm 0cm 0.1cm},clip]{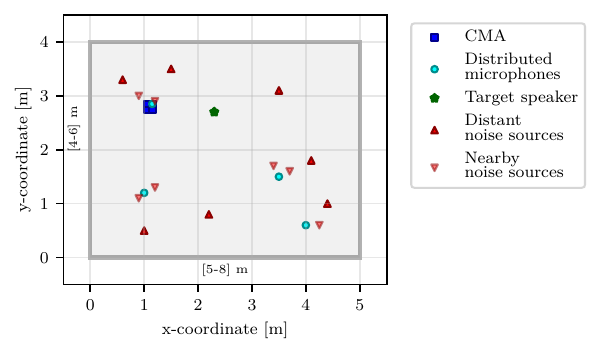} %
    \vspace{-0.4cm}
    \caption{Exemplary evaluation scene with a compact microphone array and distributed microphones, and different noise conditions (distant and nearby).}
    \label{fig:scene}
\end{figure}
\noindent
For the evaluation, we generated room impulse responses using pyroomacoustics \cite{Scheibler2018pyroomacoustics}, for which an exemplary scene is shown in Fig. \ref{fig:scene}. We simulated a total of five different rooms with dimensions between [5-8]$\times$[4-6]$\times$[2.5-3.5] m, reverberation times between 200  ms and 700 ms and different source and microphone positions. 
Each scene contained a single target speaker and seven noise sources. 
We considered two different microphone configurations to investigate the effects of multi-channel
masking: 1) a square four-channel compact microphone array (CMA) with a length and width of 4 cm (dark blue in Fig. \ref{fig:scene}) and 2) distributed microphones (light blue) consisting of four microphones (note that the first microphone of the CMA was also used for the distributed microphones). The CMA, containing the common reference microphone for both configurations, was placed closest to the target source.
All sources and microphones were placed at a height of 1.5 m and had a minimum distance of 0.5 m to the walls. We considered two different scenarios for the noise positions: In the first scenario, all noise sources were placed at a minimum distance of 0.5 m to all microphones and all other sources, referred to as distant sources (dark red in Fig. \ref{fig:scene}). In the second scenario, we considered nearby noise sources (light red in Fig. \ref{fig:scene}), where noise sources were placed a maximum of 0.5 m away from the nearest microphone to assess the effect of highly disjoint noise signals across microphones for different masking approaches.

As target speech, ten utterances from the WSJ0 corpus \cite{WSJ0} were used, while the nonstationary noise signals were taken from the DNS3 corpus \cite{DNS3} with a random gain in a range of [-10,10] dB applied to the source signals. 
Speech and noise were mixed at an input SNR of\; \{0,5,10\} dB in the reference microphone, and white noise was added at an SNR of 40 dB to simulate sensor noise.
Note that although the scenarios were spatially stationary, the intermittency of the different sources caused temporal changes, especially in the noise covariance matrix.

All signals were generated and processed at a sampling frequency $f_{\rm s}$ = 16 kHz.
As evaluation metrics, we used the SNR and PESQ improvement ($\Delta$SNR and $\Delta$PESQ) \cite{Rix2001PESQ} relative to the input at the reference microphone signal. All metrics were computed on the respective time-domain signals.

\vspace{-0.2cm}
\subsection{Implementation and Algorithmic Settings}\label{subsec:algo}

\begin{figure}
    \centering
    \includegraphics[width=0.99\linewidth,trim={1cm 0cm 1cm 0cm},clip]{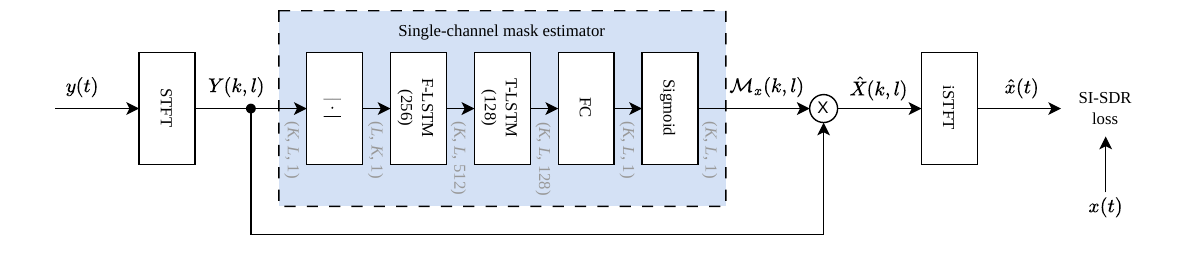}
    \vspace{-0.5cm}
    \caption{DNN-based single-channel mask estimator depicting the training pipeline using an SI-SDR loss. The blue block indicates the mask estimator used for inference. Feature dimensions are indicated in gray.}
    \vspace{-0.1cm}
    \label{fig:MaskDNN}
\end{figure}
For the implementation of the mask-based MVDR beamformer in \eqref{eq:w2}, we used an STFT with a frame length $N_{\rm FFT}$ = 512 samples ($K$ = 257 frequency bins), an overlap of 50\% ($N_{\rm overlap}$ = 256) and a square-root-Hann window for analysis and synthesis.
To estimate the covariance matrices, we considered either the batch implementation in \eqref{eq:batchCovMatVec} by averaging over an entire utterance, or the online implementation in \eqref{eq:onlineCovMat} with temporal contexts of $\tau = \{1000,500,200\}$ ms, where the number of past frames is obtained as $P = \lceil (\tau f_{\rm s}-N_{\rm FFT}) / (N_{\rm FFT}-N_{\rm overlap}) \rceil$, where $\lceil\cdot\rceil$ is the ceil operator. All covariance matrices were regularized with a scaled identity matrix, i.e., $ \Rest_{\nu,\mathrm{reg}} = \Rest_{\nu} + \delta \mathbf{I}$ with $\delta=10^{-10}$.

As indicated in Section \ref{subsec:mask}, we used three different mask vectors (ref, mean and multi). We further considered two methods for the estimation of the speech mask $\mask_x$: 

1) An \textbf{ideal ratio mask (IRM)} that serves as an optimal reference, computed as the oracle Wiener gain, i.e., 
\begin{equation}\label{eq:WienerGain}
    \mask_{x,m}(k,l) = \frac{|X_m(k,l)|^2}{|X_m(k,l)|^2 + |N_m(k,l)|^2}\, .
\end{equation}

2) A \textbf{DNN-based mask estimator} to assess the performance of different masking approaches for a realistic online mask estimator. The considered DNN architecture, shown in Fig. \ref{fig:MaskDNN}, is inspired by the architecture proposed in \cite{Tesch2022fixedJNF}. We considered the absolute value of a single microphone signal in the STFT domain as the input feature to predict a real-valued mask.
The mask estimator consisted of three trainable layers: a bi-directional LSTM-layer across the frequencies (F-LSTM) with 256 units, a uni-directional LSTM across the time-frames (T-LSTM) with 128 units allowing for frame-causal processing, and a fully connected (FC) layer with sigmoid activation.
The DNN was trained end-to-end with the scale-invariant signal-to-distortion ratio SI-SDR$(x(t),\;\hat{x}(t))$ \cite{SISDR} as loss function, i.e., as a single-channel speech enhancement system independent of the mask-based beamformer. We used the Adam optimizer \cite{Kingma2015Adam} in its standard configuration with a learning rate of $10^{-3}$ and a batch size of 8 utterances of 4 s length.
The training dataset comprised 30 hours of training and one hour of validation data with speech and noise material from the same corpora as for the evaluation, at input SNRs in the range of [-10, 20] dB, with no overlap of training and evaluation data. To account for room effects, the clean signals were convolved with RIRs similar to the ones used in the evaluation.
The training was terminated after a maximum of 50 epochs with early stopping if no improvement was observed in the validation loss for five epochs. 

Similarly to \cite{zhang2021AdHoc,jukic2023TAC,Tesch2022fixedJNF}, the noise mask was obtained complementarily to the speech mask, i.e., $\mask_{n,m}(k,l) = 1- \mask_{x,m}(k,l)$.
Note that $\mask_n$ can generally also be estimated independently of $\mask_x$.
All masks were constrained to the range $[0.01,1/(1+\varepsilon)]$ with $\varepsilon = 10^{-5}$.

\vspace{-0.2cm}
\subsection{Results}\label{subsec:Results}

\subsubsection{Temporal Contexts}\label{subsubsec:context}
\begin{figure}[t]
    \centering
    \includegraphics[width=0.95\linewidth,trim={0cm 0.225cm 0cm 0.2cm},clip]{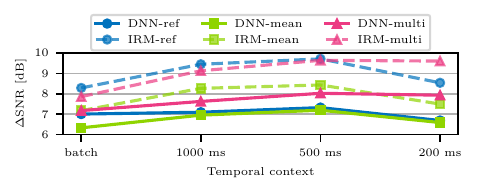} 
    \vspace{-0.3cm}
    \caption{Average SNR improvement for different temporal contexts and masking approaches using distributed microphones.}
    \label{fig:ResultsContext}
\end{figure}

\noindent
In this experiment, we investigate the effect of different temporal contexts on the performance of mask-based beamformers in distributed microphones with intermittent noise sources.
Fig. \ref{fig:ResultsContext} shows average SNR improvement using the four distributed microphones, plotted over different temporal contexts. In this experiment, we use both the IRMs and the DNN-based masks.
The shown results are averaged over rooms, utterances, and SNRs, i.e., a total of 150 scenarios.

Firstly, it can be observed that for all three masking approaches (ref, mean, and multi) and both types of masks (IRM and DNN), a temporal context of 500 ms yields the best performance in terms of SNR improvement. This result implies a benefit of an online implementation for scenarios with nonstationary/intermittent sources. 

When comparing the IRM to the DNN-based masks, it can be observed that generally, as expected, the oracle IRM leads to higher SNR improvements than the blind DNN-based mask estimator. Interestingly, the single IRM corresponding to the reference microphone (IRM-ref) performs similarly to the multi-channel masking (IRM-multi), which implies that a well-chosen reference mask generally can perform equally well as the multi-channel masking approach, while an averaged mask (IRM-mean) is constantly outperformed by the multi-channel mask.
On the other hand, for the blind DNN-based mask estimator, the multi-channel masking approach constantly outperforms the single masks, implying a benefit of using multi-channel masking for realistic mask estimators.



\subsubsection{Microphone Configurations}\label{subsubsec:Mics}
\begin{figure}[t]
    \centering
    \includegraphics[width=0.95\linewidth,trim={0cm 0.1cm 0cm 0.2cm},clip]{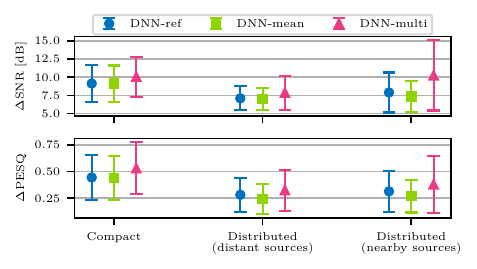}
    \vspace{-0.5cm}
    \caption{Average SNR and PESQ improvements for different microphone configurations and masking approaches.}
    \label{fig:ResultsScenes}
\end{figure}

For the two different microphone configurations (CMA and distributed microphones with distant and nearby noise sources), Fig. \ref{fig:ResultsScenes} shows the SNR and PESQ improvements for the different masking approaches (using an online implementation with 500 ms temporal context and the DNN-based mask estimator).

For the CMA, all three masking approaches perform very similarly, with a slightly better performance of the multi-channel masking approach. This result is as expected, since all masks -- and therefore also the resulting mask vectors -- are rather similar.
For the distributed microphones with noise sources spread across the room (distant sources), the differences across different masking approaches are also rather low, implying that even for distributed microphones, masks do not necessarily vary drastically across microphones. 
However, for the distributed microphones with nearby noise sources, the microphone signals -- and hence also the masks -- vary more drastically across microphones, showing a clear benefit of the multi-channel masking approach over the single masks (ref and mean).
microphones than for the CMA.


\vspace{-0.1cm}

\subsubsection{Influence of Input SNR}
\begin{figure}[t]
    \centering
    \includegraphics[width=0.95\linewidth,trim={0cm 0.1cm 0cm 0.2cm},clip]{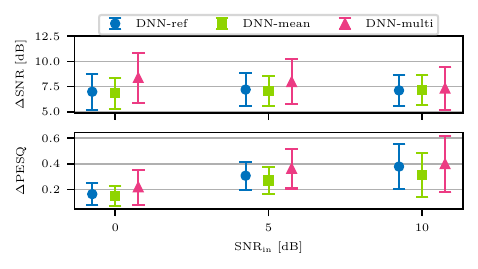}
    \vspace{-0.5cm}
    \caption{Average SNR and PESQ improvements for different input SNRs and masking approaches using distributed microphones.}
    \label{fig:ResultsSNRs}
\end{figure}

In this experiment, we investigate the robustness of the different masking approaches at different input SNRs for distributed microphones with distant noise sources. Accordingly, the results shown in Fig. \ref{fig:ResultsSNRs} are only averaged over rooms and utterances, i.e., 50 scenarios per SNR condition.
It can be observed that at high SNRs, where all microphones capture the target speech reliably, all masking approaches perform similarly in terms of SNR and PESQ improvements.
However, at low SNRs, where the captured microphone signals differ more substantially across microphones and where the mask estimator is more susceptible to errors, the differences between masking approaches become more pronounced, leading to a better performance of the multi-channel masking approach.
This implies a higher robustness of multi-channel masking over using a single mask for covariance estimation in adverse scenarios where mask estimation is subject to estimation errors.

\vspace{-0.1cm}
\section{Conclusions}\label{sec:Conclusions}

This work demonstrated the effect of extending mask-based beamforming to a multi-channel pre-filtering formulation in an application with distributed microphones. While compact arrays see little difference between single- and multi-channel masking, distributed microphones can benefit from per-channel masking, particularly when capturing different noise sources. The online implementation further highlights that short temporal contexts allow multi-channel masks to adapt effectively to spectro-temporal nonstationarity. 
A comparison of an ideal ratio mask with a blind DNN-based mask estimator that is susceptible to errors suggests higher robustness of the multi-channel masking approach compared to using single masks in realistic applications. 
Taken together, these findings suggest that accounting for spatial diversity through multi-channel masking can improve the applicability of beamforming in  geometry-agnostic, distributed microphone configurations.

{
\bibliographystyle{IEEEbib}
\bibliography{mybib}}

\end{document}